\documentstyle[epsf,epic]{ioplppt}     

\begin{document}

\title{(2,2)-Formalism of General Relativity:
An Exact Solution}
\author
{ J.H. Yoon\ftnote{3}{e-mail address:yoonjh@cosmic.konkuk.ac.kr}}

\address{
Department of Physics and Institute for Advanced Physics\\
Konkuk University, Seoul 143-701, Korea}

\begin{abstract}
We discuss the (2,2)-formalism of general relativity based on 
the (2,2)-fibration of a generic 4-dimensional spacetime 
of the Lorentzian signature.
In this formalism general relativity is describable 
as a Yang-Mills gauge theory defined on the (1+1)-dimensional 
base manifold, whose local gauge symmetry is the group of 
the diffeomorphisms of the 2-dimensional fibre manifold. 
After presenting the Einstein's field equations in this formalism,
we solve them for spherically symmetric case 
to obtain the Schwarzschild solution.
Then we discuss possible applications of this formalism.

\end{abstract}


\section{Introduction}
There have been considerable efforts and some success to reformulate 
general relativity as a gauge theory\cite{ashtekar86,ashtekar87}. 
The major advantages of such a gauge theory description of general 
relativity are, first of all, that it allows us to understand 
general relativity in terms of familiar notions of gauge 
theories, and second, that the gauge constraints associated 
with the gauge symmetry become relatively easy to handle.
However, it seems that 
a gauge theory formulation of general relativity 
as a {\it second-order}
system of partial differential equations of 
standard gauge fields as in a Yang-Mills theory 
is still missing. 
Such a description would put the spacetime physics 
into a new perspective, and hopefully, it might shed light on 
some unsolved issues related to the problem of constraints, 
in particular.

Recently, we have proposed 
a (2,2)-formalism\cite{yoon92,yoon93a,yoon96} 
of general relativity based on the 
(2,2)-fibration\cite{dinverno-stachel78,hayward93} of a generic 
4-dimensional spacetime of the Lorentzian signature. 
In this formalism the 4-dimensional spacetime manifold 
is regarded as a local product
of a (1+1)-dimensional base manifold of the Lorentzian signature 
and a 2-dimensional spacelike fibre manifold. 
By introducing {\it Yang-Mills connections} adapted 
to this fibration, it was shown that
general relativity of 4-dimensional spacetimes can be written
as a Yang-Mills gauge theory defined on the (1+1)-dimensional 
base manifold, whose local gauge symmetry is
the group of the diffeomorphisms of the 2-dimensional fibre
space. The appearance of the diffeomorphisms of the 2-dimensional 
fibre space as the Yang-Mills gauge symmetry, among others, 
is the most distinguishable feature 
of this formalism, which is valid for a {\it generic} spacetime 
that does not have any isometries.

In this paper, after a brief introduction to 
the general formalism\cite{yoon96} which will fix the 
notations, we shall present the Einstein's field equations 
written in terms of the gauge theory variables.
As an application of the (2,2)-formalism, we shall 
solve the field equations for spherically symmetric case
to find the Schwarzschild solution.
Then we discuss a few applications of this formalism.

\section{Kinematics}

Let us start by recalling that any 4-dimensional manifold
of the Lorentzian signature 
can be viewed as a local product of the (1+1)-dimensional
base manifold  $M_{1+1}$ of the Lorentzian signature, 
which is generated by 
$\partial / \partial x^{\mu} (=\partial_{\mu};\mu=+,-)$, 
and a 2-dimensional spacelike fibre manifold $N_{2}$, which is
generated by 
$\partial /  \partial y^{a} (=\partial_{a};a=2,3)$ (see Figure 1).  
It is convenient to introduce the {\it horizontal lift vector fields}
$\hat{\partial}_{\mu}$ orthogonal to 
$\partial /  \partial y^{a}$, which in general can be written as 
linear combinations of 
$\partial_{\mu}$ and $\partial_{a}$,
\begin{equation}
\hat{\partial}_{\mu}:=\partial_{\mu}
- A_{\mu}^{\ a}\partial_{a},  \label{horizontal}
\end{equation}
where $-A_{\mu}^{\ a}$ are the coefficient {\it functions}
of the linear combinations.
The fields $A_{\mu}^{\ a}$ play the role of 
{\it gauge connections} whose value lies in the Lie algebra of the
diffeomorphisms of $N_{2}$.
Let the inverse metric of the horizontal space spanned by
the horizontal lift vector fields $\hat{\partial}_{\mu}$
be $\gamma^{\mu\nu}$, and the inverse metric 
on the fibre space $N_{2}$ be $\phi^{a b}$, respectively. 
Then, in the horizontal lift basis that consists of 
$\{ \hat{\partial}_{\mu}, \partial_{a} \}$,
the inverse metric of the 4-dimensional spacetime in this 
(2,2)-fibration can be written as
\begin{equation}
\Big( {\partial \over \partial s}\Big)^{2}=\gamma^{\mu\nu}
\Big( \partial_{\mu} - A_{\mu}^{\ a}\partial_{a} \Big)
\otimes
\Big( \partial_{\nu} - A_{\nu}^{\ b}\partial_{b} \Big)
+\phi^{a b}\partial_{a}\otimes \partial_{b}.  \label{dualmetric}
\end{equation}
In the corresponding dual basis
$\{ dx^{\mu}, dy^{a} + A_{\mu}^{\ a}dx^{\mu} \}$,
which preserves the orthogonality of the decomposition,
the metric of the spacetime becomes 
\begin{equation}
ds^{2}=\gamma_{\mu\nu}dx^{\mu}dx^{\nu} + \phi_{a b}
\Big( dy^{a} + A_{\mu}^{\ a} dx^{\mu} \Big)
\Big( dy^{b} + A_{\nu}^{\ b} dx^{\nu} \Big).  \label{metric}
\end{equation}
Using the gauge freedoms associated with the spacetime 
diffeomorphisms, we can always gauge fix the horizontal 
space metric $\gamma_{\mu\nu}$ to the following form
\begin{eqnarray}
\gamma_{\mu\nu}=
\left( \begin{array}{rr}
                       -2h & -1 \\
                       -1  &  0 
       \end{array}\right).      \label{gaugefix}
\end{eqnarray}
Let us denote the new coordinates 
$\{x'^{+},x'^{-},y'^{a}\}$
in which $\gamma_{\mu\nu}$ assumes the above form as 
\begin{equation}
x'^{+}=u,  \hspace{.5cm}
x'^{-}=v,  \hspace{.5cm}
y'^{a}=y^{a},
\end{equation}
and decompose the metric $\phi_{ab}$ 
on the fibre space $N_{2}$ as 
\begin{equation}
\phi_{ab}={\rm e}^{\sigma} \rho_{ab},
\end{equation}
where 
\begin{equation}
{\rm det}\ \rho_{ab}=1.                        \label{det}
\end{equation}
The field $\sigma$ is a measure of the area element and 
$\rho_{ab}$ is the conformal 2-geometry of $N_{2}$, respectively.
Then the metric (\ref{metric}) becomes 
\begin{equation}
\fl
ds^2 
= -2dudv - 2hdu^2 +{\rm e}^{\sigma} \rho_{ab}
    \left( dy^a + A_{+}^{\ a}du +A_{-}^{\ a} dv \right)
   \left( dy^b + A_{+}^{\ b}du +A_{-}^{\ b} dv \right). \label{yoon}
\end{equation}
The spacetime geometry of the metric (\ref{yoon}) 
can be also understood as follows. 
The surface $u={\rm constant}$ defines a null hypersurface, 
and the surface $v={\rm constant}$ is either a timelike, null, 
or spacelike hypersurface, depending on the signature of $2h$.
The intersection of two hypersurfaces $u,v={\rm constant}$
is the spacelike two-surface $N_{2}$.
Notice that the horizontal lift vector fields 
$\{ \hat{\partial}_{+}, \hat{\partial}_{-} \}$ 
become
\begin{eqnarray}
& \hat{\partial}_{+}=
\partial_{+}-A_{+}^{\ a}\partial_{a},   \label{plushorizon}\\
& \hat{\partial}_{-}=
\partial_{-}-A_{-}^{\ a}\partial_{a},   \label{minushorizon}
\end{eqnarray}
but now $\hat{\partial}_{+}$ has a norm  $-2h$, and
$\hat{\partial}_{-}$ is a null vector field 
since its norm is zero. By a suitable rescaling,
the inner product of $\hat{\partial}_{+}$ and 
$\hat{\partial}_{-}$ is normalized to $-1$, 
which makes $v$ an affine parameter. 
Thus two metric functions are gauged away in (\ref{yoon}),
and the remaining eight fields 
$\{ h, \sigma, \rho_{ab}, A_{\pm}^{\ a} \}$ 
are functions of all the coordinates ($u,v,y^{a}$), since we assume no
isometries. If we further introduce the condition that 
\begin{equation}
A_{-}^{\ a}=0,
\end{equation}
then the above metric
becomes identical to the Newman-Unti metric\cite{newman-unti62}. 
In this paper, however, we shall retain the $A_{-}^{\ a}$ field, 
since its presence will make the gauge theory aspect of the formalism 
transparent. 

The integral of the scalar curvature 
of the spacetime described by the metric (\ref{yoon})
is given by 
\begin{eqnarray}
\fl
I_{\rm 4}& =   \int \! \! du \, dv \, d^{2}y \, 
   {\rm e}^{\sigma}R_{4}           \nonumber\\
\fl & = \int \! \! du \, dv \, d^{2}y \,
\Big[ 
-{1\over 2}{\rm e}^{2 \sigma}\rho_{a b}
  F_{+-}^{\ \ a}F_{+-}^{\ \ b}
  +{\rm e}^{\sigma} (D_{+}\sigma) (D_{-}\sigma) 
  -{1\over 2}{\rm e}^{\sigma}\rho^{a b}\rho^{c d}
 (D_{+}\rho_{a c})(D_{-}\rho_{b d})  \nonumber\\
\fl & +{\rm e}^{\sigma} R_2
-2{\rm e}^{\sigma}(D_{-}h)(D_{-}\sigma) 
- h {\rm e}^{\sigma}(D_{-}\sigma)^2
+{1\over 2}h  {\rm e}^{\sigma}\rho^{a b}\rho^{c d}
 (D_{-}\rho_{a c})(D_{-}\rho_{b d}) \Big]   \nonumber\\
\fl & + {\rm surface} \ {\rm terms}.   \label{barelag}
\end{eqnarray}
Here $R_{2}$ is the scalar curvature of $N_{2}$, and 
the notations are summarized below:
\begin{eqnarray}
\lo F_{+-}^{\ \ a}&=\partial_{+} A_{-} ^ { \ a}-\partial_{-}
  A_{+} ^ { \ a} - [A_{+}, A_{-}]_{\rm L}^{a}  \nonumber\\
& =\partial_{+} A_{-} ^ { \ a}-\partial_{-}
  A_{+} ^ { \ a}-A_{+}^{\ c}\partial_{c}A_{-}^{\ a}
  +A_{-}^{\ c}\partial_{c}A_{+}^{\ a},   \label{field}\\
\lo D_{\pm}\sigma & = \partial_{\pm}\sigma
-[A_{\pm}, \sigma]_{\rm L}        \nonumber\\
  & =\partial_{\pm}\sigma
-A_{\pm}^{\ a}\partial_{a}\sigma
-\partial_{a}A_{\pm}^{\ a},         \label{get}\\
\lo  D_{\pm}h & = \partial_{\pm}h - [A_{\pm}, h]_{\rm L} \nonumber\\
  & =\partial_{\pm}h -A_{\pm}^{\ a}\partial_{a}h, \nonumber\\
\lo  D_{\pm}\rho_{a b} & =\partial_{\pm}\rho_{a b}
   - [A_{\pm}, \rho]_{{\rm L}a b}       \nonumber\\
  & =\partial_{\pm}\rho_{a b}
-A_{\pm} ^ { \ c}\partial_c \rho_{a b}
-(\partial_a A_{\pm} ^ { \ c})\rho_{c b}
-(\partial_b A_{\pm} ^ { \ c})\rho_{a c}
+(\partial_c A_{\pm} ^ { \ c})\rho_{a b},    \label{rhod}
\end{eqnarray}
where $[A_{\pm}, \ast]$ is the Lie derivative 
of $\ast$ along the vector field
$A_{\pm}:=A_{\pm}^{\ a}\partial_{a}$.
Each term in $I_{4}$ strongly suggests that
the integral should be interpreted as an action integral of 
a Yang-Mills type gauge theory defined on the (1+1)-dimensional
base manifold $M_{1+1}$, interacting with 
$h$, $\sigma$, and $\rho_{a b}$.
The associated local gauge symmetry is 
the diff$N_{2}$ symmetry,
the group of the diffeomorphisms of the fibre space $N_{2}$,
which is built-in into the theory via the Lie derivatives.
It must be mentioned here that each term in (\ref{barelag}) is
manifestly diff$N_{2}$-invariant, and that the $y^{a}$-dependence 
is completely hidden in the Lie derivatives.
In this sense we may regard the fibre space $N_{2}$ as 
a kind of an {\it internal} space
as in Yang-Mills theories\cite{foot}.

\section{The Einstein's equations}

By varying $I_{4}$ with respect to the eight fields 
$h$, $A_{\pm}^{\ a}$, $\sigma$, 
and $\rho_{ab}$ (subject to the condition ${\rm det}\ \rho_{ab}=1$), 
one can obtain 
the eight field equations (eqs. (\ref{cc}), $\cdots$, (\ref{gg})) 
in the gauge (\ref{gaugefix}). 
In addition, there are two supplementary equations 
(eqs. (\ref{aa}) and (\ref{bb}) ) associated with 
the gauge fixing of the horizontal space metric 
$\gamma_{\mu\nu}$ to the form (\ref{gaugefix}).
In fact one can obtain the complete set of the Einstein's field 
equations at once by varying the Einstein-Hilbert 
action\cite{yoon96}{\it before} 
one introduces any gauge condition, 
and then imposing the gauge condition 
(\ref{gaugefix}) in the resulting field equations. 
It turns out that the eight field equations one obtains from $I_{4}$ 
by variations are identical to the eight Einstein's equations 
that follow from the general Einstein-Hilbert action
by the corresponding variations. 
The underlying reason for this equality is because 
the gauge fixing condition (\ref{gaugefix}) puts no restrictions 
on the remaining eight fields, so that the variations of 
the integral $I_{4}$ with respect to the eight fields 
are completely arbitrary variations 
subject to no restrictions at all. 
Hence the integral $I_{4}$ may be regarded an action integral, 
{\it modulo} two supplementary equations that should be appended
to it by the Lagrange multipliers, 
since it reproduces the ten Einstein's equations 
in the gauge (\ref{gaugefix}) when implemented 
by the two equations.
The complete set of the Einstein's equations are found to be
\begin{eqnarray}
\fl (a) & 
{\rm e}^{\sigma} D_{+}D_{-}\sigma 
+ {\rm e}^{\sigma} D_{-}D_{+}\sigma  
+ 2{\rm e}^{\sigma} (D_{+}\sigma)(D_{-}\sigma)
- 2{\rm e}^{\sigma}(D_{-}h)(D_{-}\sigma)  \nonumber\\
\fl     & 
- {1\over 2}{\rm e}^{ 2 \sigma}\rho_{a b}
   F_{+-}^{\ \ a}F_{+-}^{\ \ b}
- {\rm e}^{\sigma} R_{2}
- h {\rm e}^{\sigma} \Big\{
(D_{-}\sigma)^{2} 
-{1\over 2}\rho^{a b}\rho^{c d} 
 (D_{-}\rho_{a c})(D_{-}\rho_{b d})\Big\}=0, \label{aa}\\
\fl (b) & 
-{\rm e}^{\sigma} D_{+}^{2}\sigma 
- {1\over 2}{\rm e}^{\sigma}(D_{+}\sigma)^{2}
-{\rm e}^{\sigma}(D_{-}h) (D_{+}\sigma)
+{\rm e}^{\sigma}(D_{+}h)(D_{-}\sigma) \nonumber\\
\fl     & 
+2h {\rm e}^{\sigma}(D_{-}h)(D_{-}\sigma) 
+{\rm e}^{\sigma}F_{+-}^{\ \ a}\partial_{a}h
-{1\over 4}{\rm e}^{\sigma}\rho^{a b}\rho^{c d} 
 (D_{+}\rho_{a c})(D_{+}\rho_{b d})
+\partial_{a}\Big( \rho^{a b}\partial_{b}h \Big) \nonumber\\
\fl     & 
+h\Big\{ - {\rm e}^{\sigma} (D_{+}\sigma) 
  (D_{-}\sigma)
+{1\over 2}{\rm e}^{\sigma}\rho^{a b}\rho^{c d} (D_{+}\rho_{a c})
    (D_{-}\rho_{b d})
+{1\over 2}{\rm e}^{2\sigma}\rho_{a b}F_{+-}^{\ \ a}F_{+-}^{\ \ b}
+{\rm e}^{\sigma}R_{2} \Big\} \nonumber\\
\fl     & 
+h^{2}{\rm e}^{\sigma}\Big\{
(D_{-}\sigma)^{2}
-{1\over 2}\rho^{a b}\rho^{c d}
(D_{-}\rho_{a c}) (D_{-}\rho_{b d})\Big\}=0,  \label{bb}\\
\fl (c) & 
2{\rm e}^{\sigma}(D_{-}^{2}\sigma) +
{\rm e}^{\sigma} (D_{-}\sigma)^{2}  
    + {1\over 2}{\rm e}^{\sigma}\rho^{a b}\rho^{c d} (D_{-}\rho_{a c})
    (D_{-}\rho_{b d})=0,                 \label{cc}\\
\fl (d) & 
D_{-}\Big( {\rm e}^{2\sigma}
\rho_{a b}F_{+-}^{\ \ b}\Big)
- {\rm e}^{\sigma}\partial_{a}(D_{-}\sigma) 
- {1\over 2}{\rm e}^{\sigma}\rho^{b c}\rho^{d e}
    (D_{-}\rho_{b d})(\partial_{a}\rho_{c e}) 
+ \partial_{b} \Big(
{\rm e}^{ \sigma}\rho^{b c}D_{-}\rho_{a c} \Big)\nonumber\\
\fl     & 
=0,                   \label{dd}\\
\fl (e) & 
-D_{+}\Big( {\rm e}^{2\sigma}
\rho_{a b}F_{+-}^{\ \ b}\Big) 
-{\rm e}^{\sigma}\partial_{a} (D_{+}\sigma )
   -{1\over 2}{\rm e}^{\sigma}\rho^{b c}\rho^{d e}
   (D_{+}\rho_{b d})(\partial_{a}\rho_{c e})  \nonumber\\
\fl & 
+\partial_{b}\Big(  
{\rm e}^{ \sigma}\rho^{b c}D_{+}\rho_{a c} \Big)
+2h{\rm e}^{\sigma}\partial_{a}(D_{-}\sigma)   
+h{\rm e}^{\sigma}\rho^{b c}\rho^{d e}
   (D_{-}\rho_{b d})(\partial_{a}\rho_{c e}) 
+2{\rm e}^{\sigma}\partial_{a}(D_{-}h)    \nonumber\\
\fl & 
-2\partial_{b}\Big(h  
{\rm e}^{ \sigma}\rho^{b c}D_{-}\rho_{a c} \Big)
=0,                              \label{ee}\\
\fl (f) & 
-2 {\rm e}^{ \sigma}D_{-}^{2} h 
-2{\rm e}^{ \sigma} (D_{-}h)(D_{-}\sigma) 
+ {\rm e}^{ \sigma}D_{+}D_{-}\sigma 
+ {\rm e}^{ \sigma}D_{-}D_{+}\sigma        
+ {\rm e}^{ \sigma} (D_{+}\sigma)(D_{-}\sigma) \nonumber\\  
\fl & 
+ {1\over 2}{\rm e}^{ \sigma}\rho^{a b}\rho^{c d} 
  (D_{+}\rho_{a c})(D_{-}\rho_{b d})   
+ {\rm e}^{2 \sigma}\rho_{a b}
    F_{+-}^{\ \ a}F_{+-}^{\ \ b}
 -2h{\rm e}^{ \sigma} \Big\{
   D_{-}^{2} \sigma +{1\over 2}(D_{-}\sigma)^{2} \nonumber\\
\fl & 
+{1\over 4}\rho^{a b}\rho^{c d}
   (D_{-}\rho_{a c})(D_{-}\rho_{b d})\Big\}=0,  \label{ff}\\
\fl (g) & 
h\Big\{ {\rm e}^{\sigma} D_{-}^{2} \rho_{ab} 
- {\rm e}^{\sigma}\rho^{c d}(D_{-}\rho_{a c})(D_{-}\rho_{b d}) 
+{\rm e}^{\sigma}(D_{-}\rho_{a b})(D_{-}\sigma) \Big\}  \nonumber\\
\fl & 
-{1\over 2}{\rm e}^{\sigma} \Big( 
D_{+}D_{-}\rho_{a b} + D_{-}D_{+}\rho_{a b} \Big) 
+{1\over 2}{\rm e}^{\sigma} \rho^{c d}\Big\{ 
(D_{-}\rho_{a c})(D_{+}\rho_{b d}) 
+(D_{-}\rho_{b c})(D_{+}\rho_{a d}) \Big\}  \nonumber\\
\fl & 
-{1\over 2}{\rm e}^{\sigma}\Big\{ 
(D_{-}\rho_{a b})(D_{+}\sigma)
+(D_{+}\rho_{a b})(D_{-}\sigma)  \Big\}  \nonumber\\
\fl & 
 +{\rm e}^{\sigma}(D_{-}\rho_{a b})(D_{-}h)                
+{1\over 2}{\rm e}^{2 \sigma}\rho_{a c}\rho_{b d}
 F_{+-}^{\ \ c}F_{+-}^{\ \ d}     
-{1\over 4}{\rm e}^{2 \sigma}\rho_{a b}
 \rho_{c d}F_{+-}^{\ \ c}F_{+-}^{\ \ d}=0. \label{gg}
\end{eqnarray}

\section{A spherically symmetric solution}

The equations (\ref{aa}), $\cdots$, (\ref{gg}) are 
the Einstein's field equations in the gauge theory variables 
in the (2,2)-fibration of a generic spacetime.
It will be instructive to find some solution of the above equations,
since it will give us an idea how to use this formalism. 
There are several classes of 
spacetimes\cite{kramer-stephani-herlt-maccallum80,griffiths91} 
to which this (2,2)-formalism is directly applicable, 
but in this paper, we are interested in solving the above equations
for spherically symmetric vacuum case. 

In order to solve the field equations in this case,
we have to construct a coordinate system adapted 
to the metric (\ref{yoon})\cite{synge60}. 
Recall that the spherical symmetry with respect 
to a given observer means 
that the metric is independent of the orientation 
at each point on the worldline $C$ 
of that observer (see Figure 2).
Let $\vartheta$ and $\varphi$ be the angular coordinates 
that define the orientation at that point. 
Then, by the spherical symmetry, 
it suffices to consider the (1+1)-dimensional subspace
defined by $\vartheta$, $\varphi={\rm constant}$. 
Let us define the coordinates $(u,v)$ of an arbitrary event 
$E$ in this subspace as follows;
(a) Given an event $E$, draw a past-directed null geodesic 
from $E$ cutting the worldline $C$ at $P$. The coordinate $v$ is
defined as the affine distance of the event $E$ from $P$ 
along the null geodesic.
(b) The coordinate $u$ measures the location of the event 
$P$ from a certain reference point $O$ along the worldline $C$.
The affine parameter $v$ has the coordinate freedom 
\begin{equation}
v \longrightarrow v'= A(u) v + B(u),     \label{aff}
\end{equation}
on each null hypersurface defined by $u={\rm constant}$.
Also there is a reparametrization invariance 
\begin{equation}
u \longrightarrow u'=f(u),
\end{equation}
where $f(u)$ is an arbitrary function of $u$.
Notice that the equation $du=0$ defines a {\it null} geodesic 
in the $(u,v)$-subspace. If we further choose 
$A=1$ and $B=0$ in (\ref{aff}), then we can 
write the metric on this subspace 
as a product of $-2du$ and $dv + h(u,v) du$, where $h$ is 
an arbitrary function of $(u,v)$. Thus the metric of 
the spherically symmetric spacetime is given by
\begin{equation}
ds^2= -2 du dv -2 h(u,v) du^{2} 
+H(u,v)(d\vartheta^2+\sin^2\vartheta d\varphi^2). \label{spher}
\end{equation}
Here the fibre space $N_{2}$ is a two-sphere $S_{2}$
of radius $H^{1/2} (H > 0)$,
whose scalar curvature is 
\begin{equation}
R_{2}=-{2\over H}.
\end{equation}
If we compare (\ref{yoon}) with (\ref{spher}), we find that 
\begin{eqnarray}
& A_{\pm}^{\ \vartheta}=A_{\pm}^{\ \varphi}=0, \nonumber\\
& \rho_{\vartheta \vartheta}
={1 \over {\rm sin} \vartheta}, \hspace{.5cm} 
\rho_{\varphi \varphi}= {\rm sin} \vartheta, \hspace{.5cm} 
\rho_{\vartheta \varphi}= 0, \nonumber\\
& {\rm e}^{\sigma}=H \ {\rm sin} \vartheta.
\end{eqnarray}
Notice that the diff$N_{2}$-covariant derivatives $D_{\pm}$
reduce to $\partial_{\pm}$ since 
$A_{\pm}^{\ \vartheta}=A_{\pm}^{\ \varphi}=0$.
Therefore the equations $(\ref{aa}), \cdots, (\ref{ff})$ become 
\begin{eqnarray}
\fl (a')&  \hspace{.5cm} \partial_{+}\partial_{-}\sigma 
 + \partial_{-}\partial_{+}\sigma  
 + 2(\partial_{+}\sigma) (\partial_{-}\sigma)
 -2 (\partial_{-}\sigma)(\partial_{-}h)
 +{2\over H} - h (\partial_{-}\sigma)^{2}=0, \label{a}\\
\fl (b')&  \hspace{.5cm} \partial^{2}_{+}{\sigma} 
 +{1\over 2} (\partial_{+}{\sigma})^{2}
 +(\partial_{+}{\sigma})(\partial_{-}h)
 -(\partial_{-}{\sigma})(\partial_{+}h)
 -2h(\partial_{-}{\sigma}) (\partial_{-}h)   \nonumber\\
\fl & 
+h \Big\{
 (\partial_{+}{\sigma})(\partial_{-}{\sigma})
 +{2\over H} \Big\}
 -h^{2}(\partial_{-}{\sigma})^{2}=0, \label{b}  \\
\fl (c')& \hspace{.5cm}2 \partial^{2}_{-}{\sigma} 
 + (\partial_{-}{\sigma})^{2}=0,  \label{c}\\
\fl (d')& \hspace{.5cm}\partial_{a}\partial_{-}
 \sigma=0,                          \label{d}\\
\fl (e')& \hspace{.5cm}\partial_{a}\partial_{+}\sigma  
 -2h \partial_{a}\partial_{-}\sigma 
 -2 \partial_{a}\partial_{-}h=0,     \label{e}\\
\fl (f')&  \hspace{.5cm}\partial_{+}\partial_{-}\sigma 
 +\partial_{-}\partial_{+}\sigma 
 +(\partial_{+}\sigma) (\partial_{-}\sigma)
 -2 \partial^{2}_{-}h 
 -2 (\partial_{-}h)(\partial_{-}\sigma)=0,\label{f}
\end{eqnarray}
respectively, and eq. (\ref{gg}) turns out to be trivial.
Let us integrate eq. (\ref{c}) first. 
It can be written as
\begin{eqnarray}
& 2\partial_{-}X + X^{2}=0,    \label{ex}\\
& X:=\partial_{-}{\sigma}.      \label{xdefine}
\end{eqnarray}
Solving eq. (\ref{ex}), we find that
\begin{equation}
X={2\over v + 2F},        \label{second}
\end{equation}
where $F$ is an arbitrary function
of $(u, \vartheta, \varphi)$. Therefore $\sigma$ becomes 
\begin{eqnarray}
\sigma &= 2{\rm ln}\ (v + 2 F)
     + G              \nonumber\\ 
& 
={\rm ln} \ H + {\rm ln}\ {\rm sin}\vartheta,
\end{eqnarray}
where $G$ is another arbitrary function of $(u, \vartheta, \varphi)$.
Since the spacetime must be asymptotically flat,
we have to choose the integral constants $F=0$ and  
$G={\rm ln}\ {\rm sin}\vartheta$. Then $\sigma$ and $H$ becomes
\begin{eqnarray}
\sigma & =2 {\rm ln}\  v + {\rm ln}\ {\rm sin}\vartheta, \nonumber\\
H &=v^{2},                              \label{easy}
\end{eqnarray}
from which it follows that
\begin{equation}
\partial_{-}{\sigma} = {2\over v}, \hspace{.5cm} 
\partial_{+}{\sigma}=0.               \label{simple}
\end{equation}
Therefore we find that eqs. (\ref{d}) and (\ref{e}) are 
trivially satisfied, and the remaining equations become
\begin{eqnarray}
\lo (a'), (b') \hspace{.5cm} & %
2\partial_{-} h
  + {2\over v}\ h - {1\over v}=0,      \label{newa} \\
\lo (f') & 
\partial^{2}_{-} h 
  + {2\over v}\ (\partial_{-} h)=0.     \label{newf}
\end{eqnarray}
Notice that eqs. (\ref{a}) and (\ref{b}) reduce to 
eq. (\ref{newa}) identically, and that eq. (\ref{newf})
is also trivial since it follows from eq. (\ref{newa}) 
by taking a derivative with respect to $v$.
Therefore we need to solve eq. (\ref{newa}) only. 
Assuming the asymptotic flatness as $v\rightarrow \infty$,
we find that $h$ becomes
\begin{equation}
2h=1-{2m\over v},                     \label{eichi}
\end{equation}
where $m$ is a constant. 
Plugging (\ref{easy}) and (\ref{eichi}) into (\ref{spher}),
we find that the spherically symmetric solution of the vacuum 
Einstein's equations is given by 
\begin{equation}
ds^2= -2dudv-( 1-{2m\over v}) du^2
 + v^{2} (d\vartheta^2+\sin^2\vartheta d\varphi^2), \label{sol}
\end{equation}
which is just the Schwarzschild solution.
Notice that the metric (\ref{sol}) is independent of $u$, 
so $u$ is the Killing time, 
as implied by the Birkhoff's theorem. 

\section{Discussions}

There are several potential applications of this formalism, 
but here we list only a few of them. 
First, this formalism provides 
a natural (1+1)-dimensional framework 
for a conventional gauge theory 
description of general relativity of 4-dimensional 
Lorentzian spacetimes,
where the local gauge symmetry is diff$N_{2}$, the {\it infinite} 
dimensional group of the diffeomorphisms of the 2-dimensional 
fibre space $N_{2}$. 
This enables us to explore certain aspects of 
the theory, such as constructing physical observables for instance,
using the diff$N_{2}$-invariant quantities. Finding 
physical observables that are {\it spacetime} diffeomorphism invariant 
is an outstanding problem in classical general relativity, 
but so far none are known. It is natural to try to construct 
such observables using this formalism.

Second, the self-dual Einstein's equations have been identified
as some types of 2-dimensional field theories that have the
area-preserving diffeomorphisms as the internal gauge 
symmetry\cite{park90}.
In our formalism, since a spacetime
is viewed as a 4-dimensional fibre bundle
whose base manifold is (1+1)-dimensional,
a (1+1)-dimensional field theory description of a generic
spacetime is an automatic consequence of this formalism. 
More specifically, 4-dimensional general relativity can be 
regarded as a (1+1)-dimensional gauge theory of an infinite dimensional 
Yang-Mills gauge symmetry, as we stressed previously.
This observation leads to the expectation 
that (1+1)-dimensional field theoretic 
methods should be also applicable to 
the studies of 4-dimensional spacetime physics 
{\it without} the self-dual restriction, by treating the diff$N_{2}$ 
gauge symmetry as a kind of ``internal'' symmetry 
as in $w_{\infty}$-gravity 
theories\cite{bakas89,bergshoeff-pope-roman-shen-stelle90}.

Third, this formalism fits most naturally to the studies of 
gravitational waves, where one of the coordinates is 
usually adapted to the congruence of the out-going null vector field. 
In our formalism, one of the reasons that the integral $I_{4}$ 
of a generic spacetime is written in such a simple and suggestive
form as in (\ref{barelag}) is in fact due to our choice of
the out-going null vector field
\begin{equation}
{\partial \over \partial v}
- A_{-}^{\ a}{\partial \over \partial y^{a}}
\end{equation}
as one of the basis vector fields. 
Moreover, it is well-known that the physical degrees 
of freedom of gravitational waves reside in the
conformal 2-geometry\cite{sachs64}, which is precisely 
the non-linear sigma field $\rho_{ab}$.
Thus the study of gravitational waves is a natural problem 
in the (2,2)-formalism.

Finally, let us also mention that to examine
exact solutions of the Einstein's equations 
in the light of the gauge theory variables in this framework 
and interpret them from the (1+1)-dimensional gauge theory
perspective is an interesting problem by itself. 
For instance, the Schwarzschild solution in this paper
may be interpreted as the ``vacuum'' configuration, 
in the sense that the gauge
fields $A_{\pm}^{\ a}$ are identically zero. 
Those solutions that do not admit physical interpretations 
in a straightforward way from the spacetime point 
of view might admit sensible interpretations 
from this (1+1)-dimensional gauge theory point of view.

\ack
This work is supported in part by KOSEF (95-0702-04-01-3).

\pagebreak

\begin{figure}
\centerline{\epsfxsize=16cm\epsfbox{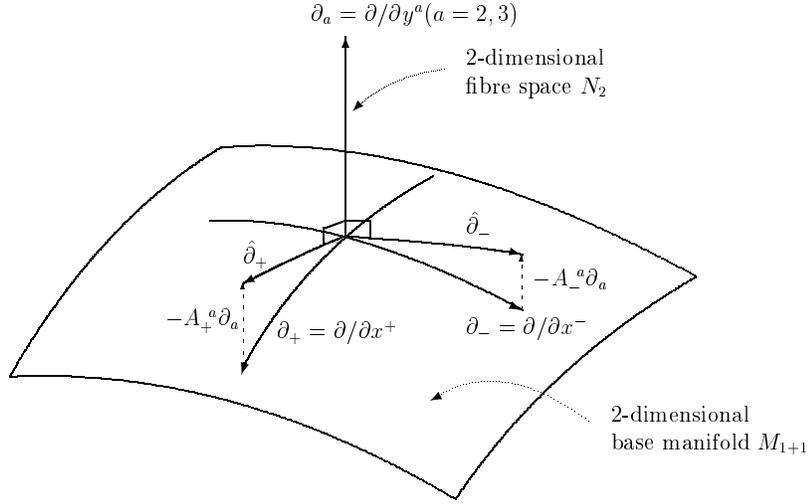}}
\vspace{-4cm}
\caption{The geometry of a 4-dimensional spacetime in the (2,2)-fibration:
the (1+1)-dimensional base manifold $M_{1+1}$ 
is generated by 
$\{ \partial/\partial x^{+}, \partial/\partial x^{-}\}$, 
and 
the 2-dimensional fibre space 
$N_{2}$ is generated by $\partial/\partial y^{a}$.
The horizontal vector fields $\hat{\partial}_{\pm}$ are 
general linear combinations of $\partial/\partial x^{\pm}$ and 
$\partial/\partial y^{a}$ with the coefficient functions
$-A_{\pm}^{\ a}$.} 
\label{exactfig1}
\end{figure}

\begin{figure}
\centerline{\epsfxsize=16cm\epsfbox{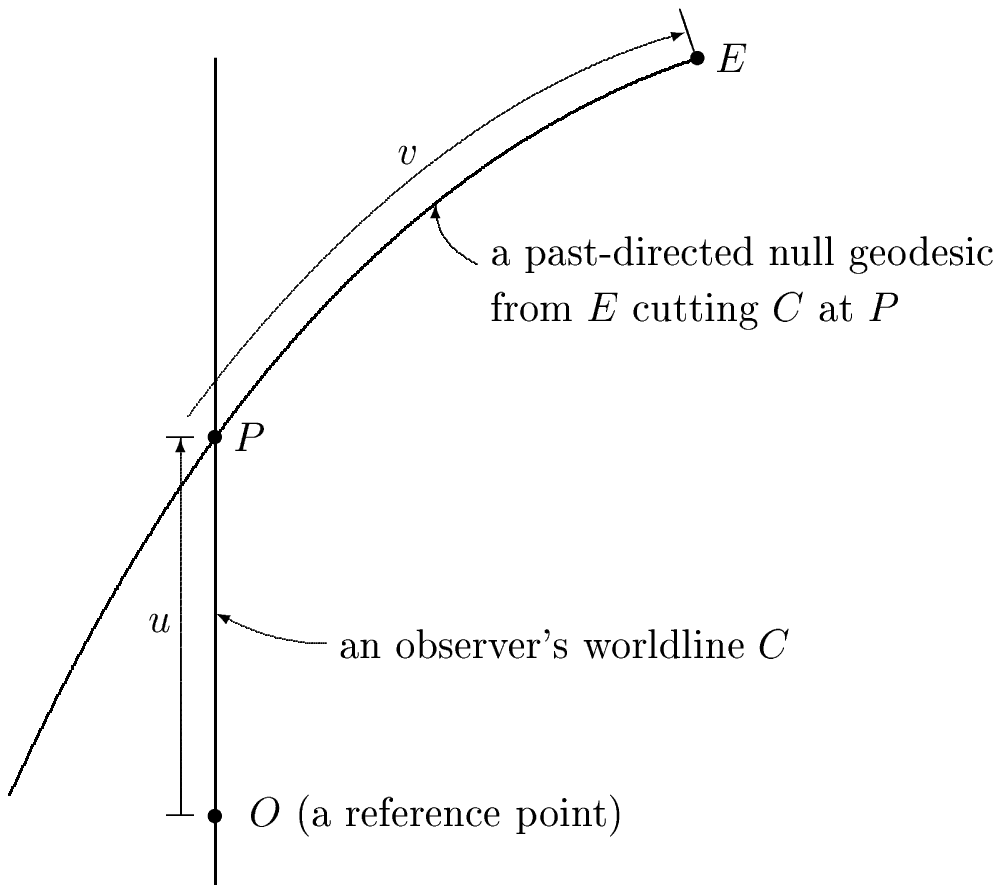}}
\vspace{-4cm}
\caption{The construction of the coordinates $(u,v)$ 
assuming the spherical symmetry about an 
observer. Here the angular coordinates ($\vartheta$, $\varphi$) are
suppressed.}
\label{exactfig2}
\end{figure}


\section*{References}
\begin{thebibliography}{99}
\bibitem{ashtekar86}{Ashtekar A 1986 {\it Phys. Rev. Lett. }
{\bf 57} 2244}
\bibitem{ashtekar87}{Ashtekar A 1987 {\it Phys. Rev.} D {\bf 36} 1587}
\bibitem{yoon92}{Cho Y M, Park Q H, Soh K S and Yoon J H 1992
  {\it Phys. Lett.} B {\bf 286} 251  }
\bibitem{yoon93a}{Yoon J H 1993 {\it Phys. Lett.} B {\bf 308} 240 }
\bibitem{yoon96}{Yoon J H 1996 {\it 4-Dimensional Kaluza-Klein 
  Approach to 
  General Relativity in the (2,2)-Decomposition of Spacetimes}, 
  gr-qc/9611050.}
\bibitem{dinverno-stachel78}{d'Inverno R and Stachel J 1978
  {\it J. Math. Phys.} {\bf 19} 2447  }
\bibitem{hayward93}{Hayward S 1993  {\it Class. Quantum Grav.}  
  {\bf 10} 779 }
\bibitem{newman-unti62}{Newman E T and Unti T 1962 
{\it J. Math. Phys.} {\bf 3} 891 }
\bibitem{foot}{Integration over the $y^{a}$-coordinates 
might be viewed as an infinite dimensional
generalization of taking finite dimensional trace
over the fibre space indices
in Yang-Mills theories with a finite dimensional gauge
symmetry}
\bibitem{kramer-stephani-herlt-maccallum80}{Kramer D, Stephani H, 
Herlt E and MacCallum M 1980 {\it Exact Solutions of Einstein's Field 
Equations} (Cambridge University Press) }
\bibitem{griffiths91}{Griffiths J B 1991 {\it Colliding Plane Waves 
in General Relativity} (Oxford University) }
\bibitem{synge60}{Synge J L 1960 {\it Relativity:The General Theory}
(North-Holland) }
\bibitem{park90}{Park Q H 1990 {\it Phys. Lett.} B {\bf 238} 287 } 
\bibitem{bakas89}{Bakas I 1989 {\it Phys. Lett.} B {\bf 228} 57 }
\bibitem{bergshoeff-pope-roman-shen-stelle90}
{Bergshoeff E, Pope C N, Roman L J, Sezgin E, Shen X 
and Stelle S 1990 {\it Phys. Lett.} B {\bf 243} 350 }
\bibitem{sachs64}{Sachs R K 1964 {\it Gravitational Radiation}, 
in Relativity, Groups
and Topology, The 1963 Les Houches Lectures, ed. deWitt B 
 and deWitt C (Gordon and Beach)}
\end{thebibliography}
\end{document}